\documentclass[AMA,STIX2COL]{WileyNJD-v2}

\articletype{Research article}%

\received{XX.XX.XXXX}
\revised{YY.YY.YYYY}
\accepted{ZZ.ZZ.ZZZZ}

\begin{document}

\title{How trace plots help interpret meta-analysis results}

\author[1]{Christian R\"{o}ver}
\author[2]{David Rindskopf}
\author[1]{Tim Friede}

\authormark{C.~R\"{O}VER, D.~RINDSKOPF, T.~FRIEDE}

\address[1]{\orgdiv{Department of Medical Statistics}, \orgname{University Medical Center G\"{o}ttingen}, \orgaddress{\state{G\"{o}ttingen}, \country{Germany}}}
\address[2]{\orgdiv{Graduate School and University Center}, \orgname{City University of New York}, \orgaddress{\state{NY}, \country{USA}}}

\corres{Christian R\"{o}ver, \email{christian.roever@med.uni-goettingen.de}}

\abstract[Abstract]{The trace plot is seldom used in meta-analysis, yet it is a very informative plot.  In this article we define and illustrate what the trace plot is, and discuss why it is important.  The Bayesian version of the plot combines the posterior density of $\tau$, the between-study standard deviation, and the shrunken estimates of the study effects as a function of $\tau$.  With a small or moderate number of studies, $\tau$ is not estimated with much precision, and parameter estimates and shrunken study effect estimates can vary widely depending on the correct value of~$\tau$.  The trace plot allows visualization of the sensitivity to~$\tau$ along with a plot that shows which values of~$\tau$ are plausible and which are implausible.  A comparable frequentist or empirical Bayes version provides similar results. 
The concepts are illustrated using examples in meta-analysis and meta-regression; 
implementaton in~\textsf{R} is  facilitated in a Bayesian or frequentist framework using the \texttt{bayesmeta} and \texttt{metafor} packages, respectively.}

\keywords{Meta-analysis, random-effects model, shrinkage, best linear unbiased prediction (BLUP).}

\jnlcitation{\cname{%
\author{C.~R\"{o}ver},
\author{D.~Rindskopf},
\author{T.~Friede}}
(\cyear{2023}), 
\ctitle{How trace plots help interpret meta-analysis results}, \cjournal{(submitted for publication)}, \cvol{2023}.}

\maketitle

\providecommand{\normaldistn}{\mathrm{Normal}}  
\providecommand{\expect}{\mathrm{E}}            

\providecommand{\notwo}{NO\textsubscript{2}}


\section{Introduction}\label{sec:intro}
  Much progress has been made in encouraging people to go beyond a fixed-effect (FE) model (also known as common-effect model) to a random-effects (RE) model in meta-analysis.\cite{BorensteinEtAl}  However, most RE models do not take into account the uncertainty in estimates of the standard deviation of true effects (commonly called tau, and symbolized as~$\tau$), and the effect that this uncertainty has on other aspects of the analysis, such as the shrunken estimates of true effect sizes.  When the number of studies is large, $\tau$~is estimated with enough precision that the effects on shrunken estimates is not worrysome; but that is not the case when the number of studies is small or even moderate (say, a dozen).\cite{FriedeRoeverWandelNeuenschwander2017a,JacksonWhite2018}

  The problem occurs with methods based on heterogeneity point estimates, such as DerSimonian and Laird, and even with empirical Bayes methods, which go part way towards a fully Bayesian solution.\cite{RaudenbushBryk1985}  On the other hand, many practitioners do not understand the Bayesian approach, in which the mathematical derivations can be taxing.  In this article we show that using graphical methods, and conceptual explanations of Bayesian models, everyone can benefit from the Bayesian approach without worrying about derivations of the results.

  A range of graphical tools have been established to aid in the interpretation or diagnosis of meta-analysis data and models, such as forest plots, funnel plots, L'Abbe plots, radial plots, or drapery plots.\cite{AnzuresCabreraHiggins2010,KossmeierTranVoracek2020,NikolakopoulouChaimani2021}
  One particular type of graphical display that (to our knowledge) was originally introduced by Rubin (1981)\citep{Rubin1981} is the \emph{trace plot}, illustrating conditional estimates over a range of plausible heterogeneity values, and by that facilitating valuable insights into the inner workings of a random-effects model, in particular the role of the heterogeneity parameter~$\tau$.
  This display seems to have been rarely picked up in the meantime; it used to be implemented in DuMouchel's ``\texttt{hblm}'' \textsc{S-plus} software package,\citep{DuMouchel1994,hblm}
  but besides some theoretical or instructional treatments\cite{RaudenbushBryk1985,GaverEtAl1992,BDA3rd,DuMouchel1994,DuMouchelNormand2000}
  we are only aware of few (published) applications of this kind of display.\cite{ZuckerEtAl1997,ShadishEtAl2013}

  The trace plot of Rubin (1981) (see supplementary Figure~\ref{fig:rubin04})\cite{Rubin1981} has been reproduced in several sources, including Gaver \emph{et~al.} (1993)\cite{GaverEtAl1992} and Gelman \emph{et~al.} (2013), although the latter has separate plots for the shrunken estimates and the posterior distribution of $\tau$.\cite{BDA3rd}  Raudenbush and Bryk (1985)\cite{RaudenbushBryk1985} included plots similar to Rubin's, and added lines for parameter estimates and a 95 percent confidence interval for those estimates.  A variation of trace plots, with the posterior density of $\tau$ plotted below the trace lines, appeared in Zucker et al (1997).\cite{ZuckerEtAl1997}

  DuMouchel (1994)\cite{DuMouchel1994} developed software for producing such plots, but it was written in \textsc{S-plus} and was not completely compatible with~\textsf{R}\@. 
  Supplementary Figure~\ref{fig:rubin03} shows a trace plot of the SAT coaching data produced by the \texttt{hblm} program of DuMouchel (see DuMouchel, 1994\cite{DuMouchel1994}). The posterior distribution of $\tau$ is plotted for $9$~values.  Because DuMouchel uses unequal spacing when picking quadrature points for $\tau$, although the height of the posterior plot bars represent probability correctly, these do not reflect the posterior \emph{density}.  Thus this plot is somewhat misleading compared to a continuous plot.  The density would be the probability divided by the bin width; bin widths in this plot are shorter for smaller values of~$\tau$.
  
  The predecessor to the trace plot is a plot that has the raw estimates on one side (sometimes left vs.\ right, sometimes top vs.\ bottom) and empirical Bayes estimates on the other side, with straight lines connecting them;
  such a display is a special case of a \emph{parallel coordinates plot}.\citep{Inselberg1985}
  These plots show only the two values of~$\tau$:  raw estimates, with $\tau=\infty$, and shrunken estimates, with $\tau$ being at its point estimate.
  An example of such a plot is in Efron and Morris (1975; p.~315).\cite{EfronMorris1975}

  We believe that the trace plot provides a very intuitive access to the inner workings of the random-effects model (also called the normal-normal hierarchical model, NNHM) that is commonly applied for meta-analysis, and may aid practicioners in the interpretation of meta-analytic inferences.  In the following, we will first introduce a motivating example in Section~\ref{sec:app:MA1-intro}, followed by a brief introduction of the random-effects model for meta-analysis in Section~\ref{sec:model}. Use of the trace plot is then extensively demonstrated using several example applications of meta-analyses and meta-regressions in Section~\ref{sec:applications}. 
  A frequentist variation of the trace plot is introduced in Section~\ref{sec:freqplot}.
  We conclude with a discussion in Section~\ref{sec:discussion}.


\section{The SAT coaching example data}\label{sec:app:MA1-intro}
    Rubin (1981)\citep{Rubin1981} discussed an application case involving what used to be called the \emph{Scholastic Aptitude Test (SAT)}; these are regularly given to support college admission decisions. While the SAT is designed to be resistant to short-term preparation exercises, this particular example dealt with the effectiveness of coaching programs to prepare students for the SAT\@. 
    The data set includes the results of eight randomized experiments (performed in eight schools), in which the adjusted effects on SAT-V (``SAT-verbal section'') scores in response to a coaching scheme were evaluated.  The effect size is the mean difference in SAT verbal scores among those coached versus those not coached in each of the eight randomized experiments.
    \begin{figure}
      \centering
      \makebox{\includegraphics[width=0.95\linewidth]{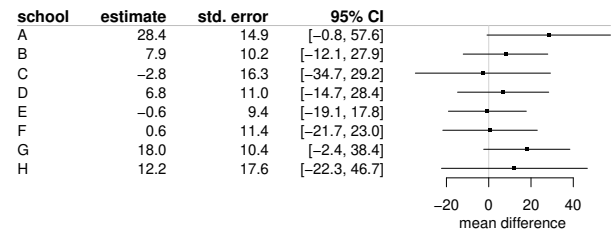}}
      \caption{\label{fig:rubin01}Forest plot for the SAT~coaching example data introduced in Section~\ref{sec:app:MA1-intro}.\citep{Rubin1981} For each of the 8~schools (labelled A--H), an estimate~$y_i$ of the coaching programme's effectiveness is given along with a standard error~$s_i$. A positive effect estimate (i.e., an observed \emph{increase} in the SAT-V score) suggests a successful programme.} 
    \end{figure}
    Figure~\ref{fig:rubin01} shows the data in a forest plot.
    Effect estimates tend to be on the positive side, suggesting successful coaching programmes; however, uncertainty is large, and none of the eight experiments was able to convincingly demonstrate effectiveness on its own.
    In addition, it is not obvious whether there are any differences between schools, i.e., whether for example the school that appeared to show the largest treatment effect (school~A) did in fact do better than the others. While the observed experimental outcomes might also be consistent with a common effect across schools, it is also conceivable that effects may vary between schools.
    A hierarchical (or \emph{random effects (RE)}) model allowing for potential heterogeneity between estimates was proposed for analysis, as described in the following section.\cite{Rubin1981}


\section{Trace plots in meta-analysis}\label{sec:model}
  \subsection{The analysis model}\label{sec:likelihood}
\begin{figure}
      \centering
      \makebox{\includegraphics[width=0.95\linewidth]{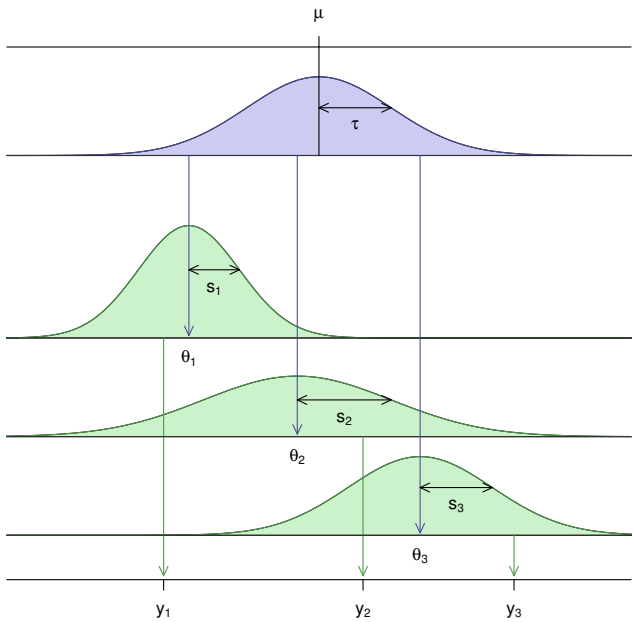}}
      \caption{\label{fig:nnhm}Illustration of the \emph{normal-normal hierarchical model (NNHM)} that is at the basis of many meta-analyses. Several studies (here: $i=1,2,3$) have slightly differing (heterogeneous) treatment effects~$\theta_i$; their underlying distribution is shown in blue at the top of the figure, and it is characterized by its center (or ``overall mean'')~$\mu$, and the degree of \mbox{(dis-)}similarity is denoted by the \emph{heterogeneity} parameter~$\tau$. When a study is conducted, we only ever get to know its true effect~$\theta_i$ with some amount of uncertainty, which is quantified through the \emph{standard error}~$s_i$. The eventual data, the effect estimates~$y_i$, are hence more or less offset from the true values~$\theta_i$, depending on the magnitudes of the~$s_i$.}
    \end{figure}

    A simple but useful and commonly applied meta-analysis approach is given through the \emph{normal-normal hierarchical model (NNHM)}, which captures the analysis problem based on normal distributions for both the measurement errors as well as for the between-study heterogeneity. The model is sketched in Figure~\ref{fig:nnhm}; a more technical treatment is provided e.g. by R\"{o}ver (2020).\cite{Roever2020}
    
    The NNHM is characterized by an \emph{overall mean} parameter~$\mu$, that denotes the ``average'' effect across all studies, and the \emph{heterogeneity}~$\tau \geq 0$, denoting the (dis-)similarity of effects in different studies. The data that are the basis for inference here are the effect estimates~$y_i$, and their associated standard errors~$s_i$ (which are treated as known). In mathematical terms, the model may be expressed as 
    \begin{eqnarray}
      \theta_i & \sim & \normaldistn(\mu,\, \tau^2)\mbox{,} \label{eqn:nnhm1}\\
      \mbox{and} \quad y_i & \sim & \normaldistn(\theta_i,\, s_i^2) 
      \quad \mbox{($i=1,\dots,k$)}\label{eqn:nnhm2}
    \end{eqnarray}
    (see also Figure~\ref{fig:nnhm}), meaning that study-specific means~$\theta_i$ may exhibit a certain amount of variation (of scale~$\tau$) around their common mean~$\mu$, while the data~$y_i$ measure the effects~$\theta_i$ only with a limited accuracy, which is expressed through the associated standard errors~$s_i$.
  
    Figure~\ref{fig:nnhm} illustrates the model graphically; equation~(\ref{eqn:nnhm1}) is represented by the blue density at the top that generates (here: three) different true effects~$\theta_i$, one for each study. According to equation~(\ref{eqn:nnhm2}), each study then yields an effect estimate~$y_i$ that tends to be some distance from~$\theta_i$, as indicated by the green densities.
    The \emph{data} eventually observed are the estimates~$y_i$ and their standard errors~$s_i$; the~$\theta_i$ as well as~$\mu$ and~$\tau$ are unknowns.
    The overall mean~$\mu$ is often the figure of primary interest; generally it may be interpreted as the \emph{average effect} across studies, in the special case of $\tau=0$, the mean~$\mu$ as well as the study-specific~$\theta_i$ all collapse to a single \emph{common effect}.
  
    In the statistical analysis, one then needs to learn the unknown model parameters (overall mean effect~$\mu$, heterogeneity~$\tau$, and study-specific effects~$\theta_i$) based on the data given in terms of the effect estimates~$y_i$ and standard errors~$s_i$.
    As one might imagine, the data may sometimes only convey a rather vague idea of the underlying true parameter values, for example in a case of only few estimates, as in the sketch in Figure~\ref{fig:nnhm}.

    Use of a normal distribution for the measurement uncertainty~(\ref{eqn:nnhm2}) may often be motivated via the central limit theorem (i.e., ``large'' sample sizes within studies); the normal distribution at the first model stage is mostly a convenient (albeit obvious) choice.
    The simple NNHM may be extended to a \emph{meta-regression} model considering study-level covariables in addition.\citep{RoeverFriede2023} For example, when each study~$i$ also provides a covariate~$x_i$, the common overall mean~$\mu$ may be replaced by an expression~$\beta_0 + \beta_1 x_i$ to account for the (potential) effect of~$x_i$ on the outcome.  With covariates in the model, $\tau$ now represents the \emph{residual} standard deviation; that is, the amount of variation not accounted for by the covariates (and standard errors).
  
  \subsection{Marginal and conditional posteriors}
    In a Bayesian context, the information extracted from a data set is formulated in terms of \emph{probability distributions}, expressing what parameter values or ranges are more or less probable \emph{given the data at hand}.\cite{BDA3rd} Depending on the problem, the data, and the assumptions implemented, these so-called \emph{posterior distributions} may convey more or less precision or ambiguity, and these may often roughly resemble the shape of a normal distribution, or may also look quite different.
    
    In the present context, we will first of all consider the posterior distributions of the heterogeneity parameter~$\tau$, and of the study specific effects~$\theta_i$ or their overall mean~$\mu$ (see Section~\ref{sec:likelihood}).
    A special variety of a posterior distribution is the so-called \emph{conditional} posterior. For example, the posterior \emph{of~$\mu$ conditional on~$\tau$} (denoted as ``$\mu|\tau$'') depends on the value of~$\tau$ that we insert on the right-hand side, and it reflects the inference on~$\mu$ if we happened to know that~$\tau$ was the actual true heterogeneity value. By varying the $\tau$~value, we may then get an impression of how our inferences depend on our information about~$\tau$. In the particular case of the NNHM, the conditional posteriors are all simply normal distributions, so that they are readily characterized through their associated (conditional) mean and variance parameters.
    
    The conditional posterior distributions have analogous (and to some extent analytically identical) counterparts in the context of frequentist/likelihood inference; a given $\tau$~value defines the \emph{conditional likelihood}, and with that, conditional maximum likelihood (ML) estimates and associated standard errors.\cite{RaudenbushBryk1985,Viechtbauer2010}

  \subsection{The trace plot}
    While the heterogeneity~$\tau$ is commonly considered a nuisance parameter, the trace plot aims to illustrate how our conclusions depend on this value, while also indicating the relevant plausible range of $\tau$~values. The $x$-axis in a trace plot corresponds to different $\tau$~values (with $\tau \geq 0$). The $y$-axis shows first of all the inferred values (conditional posterior means) of the study-specific effects~$\theta_i$, as well as model parameters such as the overall mean~$\mu$ (or regression parameters~$\beta_j$, or linear combinations of these). The trace plot is sometimes also called a \emph{conditional means} or \emph{conditional shrinkage plot}.
      
    A trace plot for the example data from Section~\ref{sec:app:MA1-intro} is shown in Figure~\ref{fig:rubin02} and is introduced and discussed in detail in the following section.


\section{Example applications}\label{sec:applications}
  \subsection{SAT coaching meta-analysis}\label{sec:app:MA1}
\begin{figure}
      \centering
      \makebox{\includegraphics[width=0.95\linewidth]{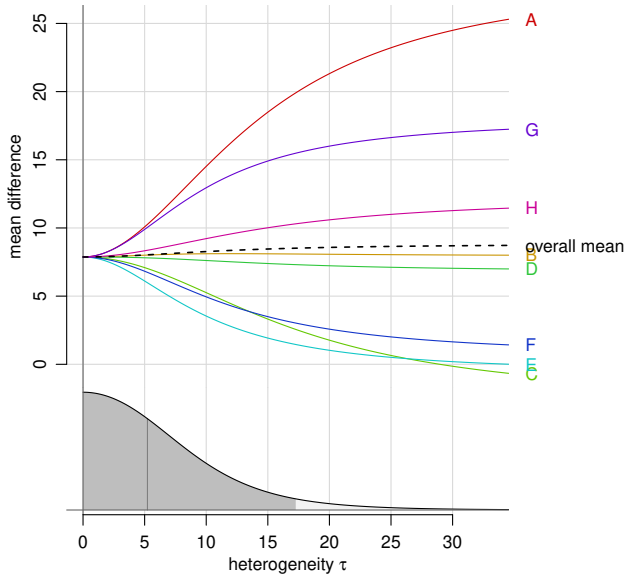}}
      \caption{\label{fig:rubin02}Trace plot for the SAT coaching example data (see Sections~\ref{sec:app:MA1-intro} and~\ref{sec:app:MA1}).\citep{Rubin1981}} 
    \end{figure}

    Figure~\ref{fig:rubin02} shows a trace plot for the SAT coaching example data  (Rubin, 1981)\cite{Rubin1981} introduced in Section~\ref{sec:app:MA1-intro}.
    Meta-analysis was performed using uniform priors for~$\mu$ and~$\tau$.  
    These ``flat''' priors are essentially uninformative and conservative specifications for both parameters.\citep{Roever2020,RoeverEtAl2021}

    At the bottom of the trace plot, the posterior distribution of the heterogeneity~$\tau$ is illustrated; its most likely value is at~$\tau=0$, but a range of positive values also remains plausible.
    The posterior median is at $5.2$, which is indicated by a vertical grey line, and the 95\% credible interval (CI), reflecting the most likely region for the true parameter value, ranges from~$0$ to~$17.3$, and is shown by the grey shaded area.
    
    The top section of the plot shows the \emph{conditional means} ($\theta_i$) for the eight schools, as well as for the overall mean~$\mu$. On the far left, the case of~$\tau=0$ corresponds to the ``homogeneous'' (or ``fixed-effect'') case where all schools' effects collapse into a common effect (of about~$7.9$).
    As soon as one considers posi\-tive $\tau$~values, the estimated (conditional) effects tend to be a compromise between the estimated overall mean~$\mu$ and the observed (empirical, apparent) effect~$y_i$. Technically, the conditional posterior means~($\expect[\theta_i|\tau]$) shown in the trace plot result as a \emph{weighted mean} of the (conditional posterior) overall mean estimate~($\expect[\mu|\tau]$) and the sample estimates~($y_i$).\citep{Roever2020,RoeverFriede2021} The attraction towards the overall mean vanishes with increasing~$\tau$.
    
    This so-called \emph{shrinkage} of individual effects towards each other (or towards their common mean) is an example of the classical \emph{``regression towards the mean''} effect.\cite{BlandAltman1994a,BlandAltman1994b,DuMouchel1994}
    This may be illustrated by considering one of the more extreme examples; consider the case of \emph{school~A} which appeared to demonstrate the greatest coaching effect (of $28.4$~points; see Figure~\ref{fig:rubin01}).
    When this datum is not considered in isolation, but in the context of the remaining observations, it appears to be a ``lucky'' outlier to some extent. Assuming $\tau=0$, the homogeneous case at the left of the trace plot, one must conclude that the observation of~$28.4$, somewhat above the probable mean of~$7.9$, was due to measurement error alone (which is not implausible given the standard error and the fact that this happened to be the maximum out of 8~measurements).\citep{Rubin1981}
    Once some positive heterogeneity is allowed for, the reasoning changes only gradually; the fact that school~A measured the greatest effect quite naturally is attributed to some degree to be due to an outlying (``lucky'') measurement, while at the same time, once differing effects between schools are permitted, school~A is likely to have the largest effect among the eight schools.

    Supplementary Figure~\ref{fig:rubin06} illustrates the ``shrinkage'' estimate for school~A in a bit more detail; in addition to the conditional mean, the conditional credible intervals (CIs) for school~A and for the overall mean~$\mu$ are shown. 
    While the shrinkage estimate for school~A is above the overall mean, the CIs are largely overlapping, which also makes sense when considering that the between-study heterogeneity~$\tau$ is estimated to most likely be smaller than the individual estimates' standard errors~$s_i$ (see Figure~\ref{fig:rubin01}).
    Consideration of such shrinkage estimates has important applications e.g. in the context of clinical trials, where a meta-analysis of ``historical'' data may contribute to the prior information considered in a new trial.\cite{RoeverFriede2020,RoeverFriede2021}
    
    The shrinkage of individual studies' effect estimates goes along with a certain precision gain; for finite $\tau$~values, the conditional standard errors are always smaller than the original $s_i$~values,\cite{WandelNeuenschwanderRoeverFriede2017} and for $\tau=0$ all estimates are completely ``pooled''.

    While the trace plot shows the \emph{conditional} means and allows for some insights into the role of the heterogeneity, eventual inference will focus on \emph{marginal} estimates (e.g., means, medians or modes of marginal distributions), i.e., consideration of (conditional) estimates integrated (marginalized) over the heterogeneity posterior distribution shown at the bottom of the trace plot. With most probability concentrated at low heterogeneity values, we may expect substantial shrinkage, and a corresponding gain in precision. 

    Supplementary Figure~\ref{fig:rubin05} shows a forest plot indicating marginal shrinkage estimates as well as the overall mean along with the data. Marginal shrinkage estimates are substantially more precise than the original data estimates, and in all cases we see a sizeable shift towards the common mean estimate.
    For example, the marginal estimate (of~$\theta_1$) for school~A is at~$10.5$ (95\% CI [$-3.3$, $29.9$]), and the associated posterior standard deviation amounts to only $56\%$~of the original~$s_1$.
    The posterior standard deviation for the overall mean~$\mu$ is at~$5.2$, and with that substantially smaller than any of the~$s_i$ provided with the data.

    As a historical digression, we include the original trace plot from Rubin (1981)\cite{Rubin1981} in supplementary Figure~\ref{fig:rubin04}. Essentially the same plot also appears in Gaver \emph{et~al.} (1992, Sec.~3.3),\citep{GaverEtAl1992}
    and the example (including slightly different plots) is also discussed in a general hierarchical modeling context in Gelman \emph{et~al.} (1995, Sec.~5.5).\cite{BDA3rd}
    Using DuMouchel's original \textsc{S-plus} code\citep{DuMouchel1994,hblm} and porting it to~\textsf{R}, we generated a trace plot for this data set, shown in supplementary Figure~\ref{fig:rubin03}.

  \subsection{Aspirin meta-analysis}\label{sec:app:MA2}
    The next example shows the utility of the trace plot for checking model assumptions and locating violations of those assumptions. 
    The data are from the widely-used meta-analysis of studies on the effect of aspirin on prevention of a second myocardial infarction (heart attack).\citep{Peto1980}
    \begin{figure}
      \centering
      \makebox{\includegraphics[width=0.95\linewidth]{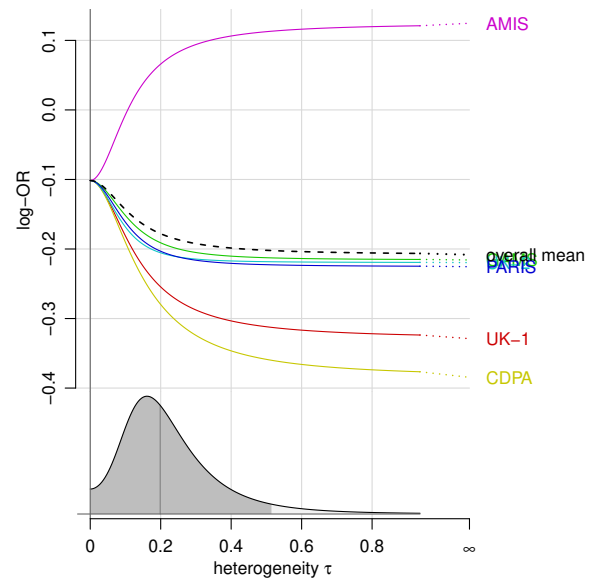}}
      \caption{\label{fig:peto01}Trace plot for the Aspirin example data.\citep{Peto1980}} 
    \end{figure}
    Figure~\ref{fig:peto01} shows a trace plot based on a simple random-effects meta-analysis of the log-odds ratios (log-ORs) for myocardial infarction when comparing aspirin to placebo.
    Note that (as in the previous examples) the trace lines level off towards large values of~$\tau$ --- the limiting values are in fact the $y_i$~values, and their arithmetic mean. In Figure~\ref{fig:peto01} the limiting values are included at the right-hand side (dotted lines) at the ``$\tau\!=\!\infty$'' $x$-axis tick mark.    
    
    At the left-hand side, for values of~$\tau$ near zero, estimates of true effect sizes are at or near the overall average, but the posterior distribution of~$\tau$ shows that zero is an unlikely value of~$\tau$.
    For values of~$\tau$ that are more likely to be true, the trace lines of shrunken estimates diverge into two parts: The first part is a group that have similar (negative) values of the effect size, and the second part is a single study (the \textsc{amis} trial) that diverges from the main group.  The single outlier not only diverges from the main group, it has an effect that is positive (suggesting that aspirin was harmful), while the others are strongly negative (see also the forest plot in supplementary Figure~\ref{fig:peto02}). 
    The heterogeneity's posterior median is at~$\tau=0.20$, and the posterior distribution largely covers $\tau$~ranges in which rather little shrinkage is taking place.
    
    The data were subsequently analyzed to investigate potential sources of heterogeneity, but with limited success; \emph{adjusted} mortality estimates seemed more homogeneous, and differences were found in short-term vs. long-term follow-up outcomes.\citep{Canner1987} A substantial fraction of the empirical heterogeneity may be attributed to the outlying \textsc{amis}~study; the posterior median for~$\tau$ is reduced to only~$0.094$ (from~$0.20$) when the \textsc{amis} study is omitted; this sensitivity analysis shows that it is the \textsc{amis} study driving the heterogeneity estimate towards larger values where less shrinkage is implied.

  \subsection{Meta-regression using binary covariables}\label{sec:app:MR1}
\begin{figure}
      \centering
      \makebox{\includegraphics[width=0.95\linewidth]{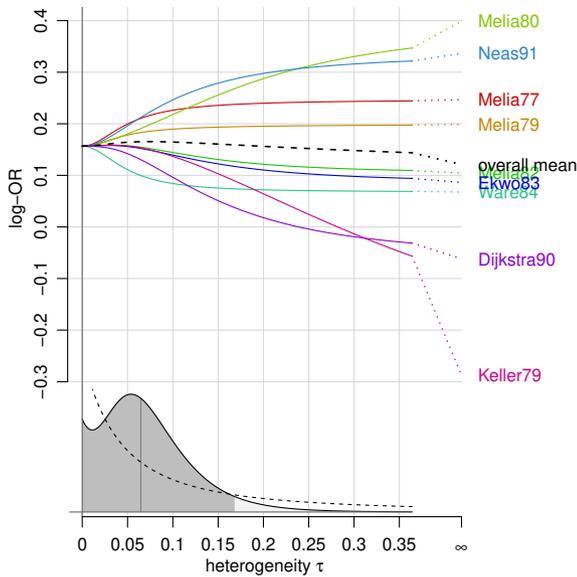}}
      \caption{\label{fig:dumouchel01}Trace plot for the {\notwo} example data.\citep{DuMouchel1994}} 
    \end{figure}

    DuMouchel (1994)\cite{DuMouchel1994} meta-analyzed nine studies that examined the relationship between nitrogen dioxide (\notwo) exposure and the development of respiratory illness in children.
    The results had originally been compiled by Hasselblad \emph{et~al.} (1992);\cite{HasselbladEddyKotchmar1992} 
    supplementary Figure~\ref{fig:dumouchel04} presents the original data as a forest plot,
    and Figure~\ref{fig:dumouchel01} shows the trace plot for a ``simple'' meta-analysis of the data.
    DuMouchel utilized a weakly informative prior distribution for the heterogeneity here;\cite{DuMouchel1994,Roever2020} 
    the prior density (dashed line) is shown along with the posterior density (solid line) at the bottom of the plot.
    In this case, one can clearly see the differing shrinkage that comes with different standard errors (see also the forest plot in supplementary Figure~\ref{fig:dumouchel04}): the study with the largest~$s_i$ (``Keller79'') has substantial shrinkage even for very large $\tau$~values, whereas the estimates with greater precision (``Ware84'', ``Melia77'', ``Melia79'') only tend to shrink towards the common overall mean for small $\tau$~values.
    Note the differing appearance when comparing the ``Keller79'' study with the case of the \textsc{amis} trial in the previous example: an outlying estimate that also has a large standard error associated has a very different effect on the analysis.
    
    The studies' one-to-one comparability, however, seemed to be questionable, and a number of distinguishing features were noted. For example, some studies reported estimates that had been adjusted for gender, while others hadn't. One may imagine that if the chances for respiratory illness differ between girls and boys, then an analysis of the {\notwo} effect may reduce bias, gain precision, and avoid confounding effects if gender is adjusted for. Such study-level covariables may be considered in a meta-regression analysis; when each study's adjustment status ($x_i=0$ if the $i$th study did adjust for gender, $x_i=1$ if it failed to adjust) is provided, one may specify a model fitting two parameters ($\beta_0 + \beta_1 x_i$) rather than a single ``intercept'' parameter~($\mu$) only.\citep{HigginsEtAl2021,RoeverFriede2023}
    \begin{figure}
      \centering
      \makebox{\includegraphics[width=0.95\linewidth]{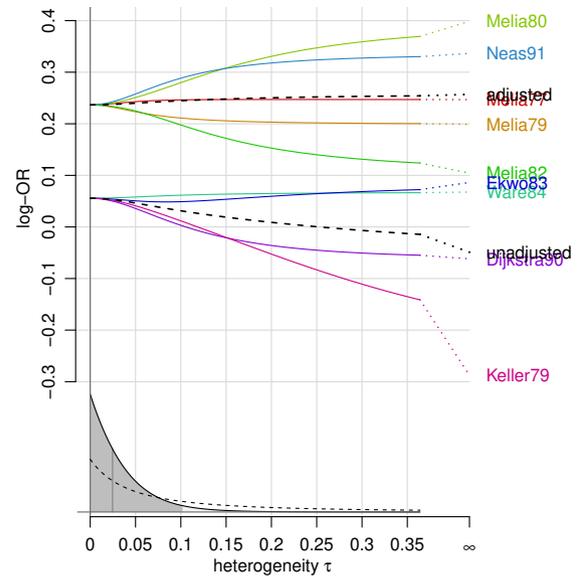}}
      \caption{\label{fig:dumouchel02}Trace plot for the {\notwo} example data; meta-regression with a single binary covariable (adjustment for gender, y/n).\citep{DuMouchel1994}} 
    \end{figure}

    Figure~\ref{fig:dumouchel02} shows the trace plot for a meta-regression considering gender adjustment as a covariable. Rather than showing the (conditional) estimates of~$\mu$ as in a ``simple'' meta-analysis, one may now include estimates of the regression parameters ($\beta_j$), or of linear combinations of these in the trace plot. With the binary coding for gender adjustment as suggested above, the estimates of $\beta_0$ and $\beta_0+\beta_1$ then correspond to the mean treatment effects of adjusted and unadjusted studies, respectively.  $\beta_1$~here corresponds to the (possible) bias due to failure to adjust for gender.
    In the trace plot, it becomes evident that the estimated heterogeneity is substantially reduced (compared to the previous analysis shown in Figure~\ref{fig:dumouchel01}); it is essentially the difference of about~$0.2$ between the two group means that is explained by the covariable and that in turn reduces the heterogeneity from a posterior median of~$0.065$ to~$0.025$.
    The individual studies' shrinkage estimates now behave quite differently; instead of converging to a common effect for zero heterogeneity at the left of the plot, each study now shrinks towards one out of the two subgroup means.
    Based on the two group's mean effects, the use of gender adjustment within a study seems to result in a larger estimated effect of {\notwo}~exposure.
    
    The importance of this meta-analysis was that it actually investigated three sources of \emph{methodological diversity}\cite{CochraneHandbook} by coding various ways that the studies could be subject to confounding, and therefore could estimate the effect size for a study that had none of those sources, even if there were no such study. There was one such study, but the others added information about the target effect, as well as information about the bias introduced by each potential source of bias.  To do this, DuMouchel fit a model with three binary dummy variables, namely \emph{gender} (whether the quoted estimate had been adjusted for gender, y/n), \emph{smoke} (whether the estimate was adjusted for patents' smoking status, y/n), and \emph{no2} (whether {\notwo}~levels were measured directly (y/n), or presence of a gas stove in the household was used as a proxy); see also the forest plot in supplementary Figure~\ref{fig:dumouchel04} for the original data.  

    These variables were all coded~1 if the study had failed to control for a condition, and~0 if it did.  Thus the intercept estimated the effect size of a study that was zero on each dummy variable, and thus had none of the three possible sources of bias.  This is similar to a suggestion of Rubin (1992)\cite{Rubin1992} that meta-analysis should be a response surface analysis, such that the main quantity estimated is the effect in an ``ideal'' study, rather than the average study.
    
\begin{figure}
      \centering
      \makebox{\includegraphics[width=0.95\linewidth]{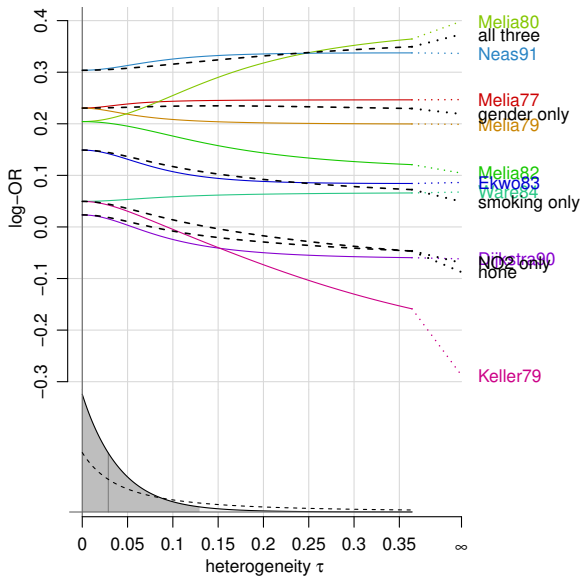}}
      \caption{\label{fig:dumouchel03}Trace plot for the {\notwo} example data,\citep{DuMouchel1994} three binary covariables (adjustment for gender, smoking, and mode of {\notwo} measurement).}
    \end{figure}

    Figure~\ref{fig:dumouchel03} shows the trace plot for the meta-regression analysis including all three covariables.
    Looking at the trace plot, one can see that the posterior distribution of~$\tau$ is highest at zero; the maximum-likelihood (ML) estimate would likewise be zero.
    Further, in an ML-approach the regression parameters would be estimated conditioning on the estimate of zero for $\tau$.  But the distribution of~$\tau$ is spread over a rather large range, meaning that moderate non-zero values are also plausible.  At those larger plausible values, the shrunken estimates would diverge some from the fully shrunken values at~$\tau=0$, but not to an extreme.
    
    Conditional estimates for five linear combinations of the regression coefficients ($\beta_j$) are also illustrated in the plot, namely, 
    corresponding to the cases where all three covariables ($x_{i1}$, $x_{i2}$, $x_{i3}$) equaled $0$ (studies with adjustment for all 3~covariables), studies without adjustment, and the three cases where one of the covariables is adjusted for.
    One can again see that each study shrinks towards an individual mean value, depending on its associated combination of covariates; those studies sharing the same combination then shrink towards a common mean.
    Consideration of covariates then allows us to investigate potential differences in effects for different study designs; it appears that a study meeting all three criteria (like the ``Neas91'' study) would yield the largest effect estimate (of about~$0.3$).  Also, a study without any adjustment has an estimate near zero, so an unadjusted study misleadingly suggests no deleterious effect.
    
    Using only nine studies to fit four regression parameters (some of which are not clearly different from zero) implies a lot of uncertainty in the model fit, which is also reflected in the heterogeneity's posterior. With a greater number of degrees of freedom in the model, the likelihood is not able to constrain the heterogeneity as much, leading to a posterior that is closer to its prior, and with that, a larger heterogeneity estimate.
    In general, one of course needs to be cautious in balancing the number of parameters estimated against the number of studies included in order to avoid overfitting; on the other hand consideration of known effect moderators may also be considered essential, and the use of informative effect priors might help here.

  \subsection{Meta-regression involving a continuous covariable}\label{sec:app:MR2}
    Karner \emph{et~al.} (2014)\cite{KarnerChongPoole2014} performed a systematic review and meta-analysis to investigate the effects of \emph{tiotropium}, a medication used in the management of \emph{chronic obstructive pulmonary disease (COPD)}.
    A total of 22~randomized placebo-controlled trials were found; among the primary endpoints was the odds ratio (OR) of exacerbation; the raw study data (with effects in terms of log-ORs) are shown in a forest plot in supplementary Figure~\ref{fig:karner05}.
    A meta-analysis of the data, using a weakly informative half-normal prior with scale~$0.5$ for the heterogeneity,\cite{RoeverEtAl2021} yields an estimated log-OR of~$-0.25$, indicating a beneficial effect of the treatment (i.e., a reduction of exacerbations).
    \begin{figure}
      \centering
      \makebox{\includegraphics[width=0.95\linewidth]{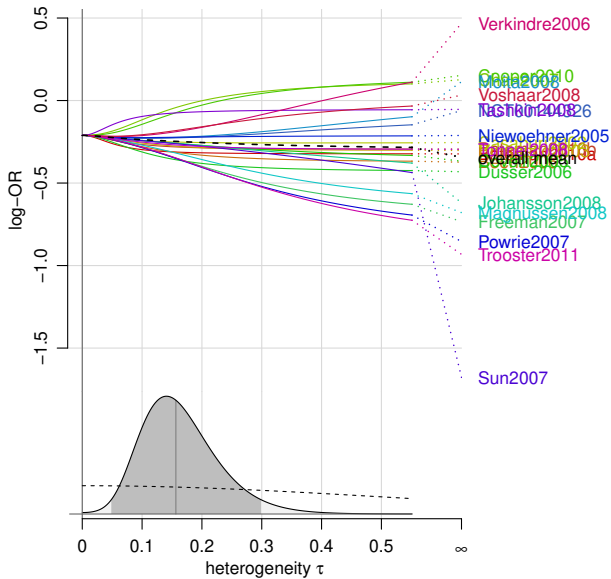}}
      \caption{\label{fig:karner01}Trace plot for the COPD example data.\citep{KarnerChongPoole2014} Meta-analysis without covariables.} 
    \end{figure}
    A trace plot for the analysis is shown in Figure~\ref{fig:karner01}.
    The half-Normal(0.5) heterogeneity prior (implying a prior median and 95\% quantile of about~$\frac{1}{3}$ and~$1$, respectively, for~$\tau$) is a conservative specification in the context of endpoints on a logarithmic scale.\cite{RoeverEtAl2021}
    The heterogeneity's posterior distribution is much narrower than the prior; it shows that $\tau$~is not likely to be zero, but also is not likely to be larger than about~$0.3$, while the prior appears effectively uniform across the relevant range.  Over the range of plausible values of~$\tau$, the shrunken effect size estimates vary, but most are in a range indicating that the drug is effective.
    Two outliers are apparent (the studies by Verkindre (2006) and Sun (2007)), which were the two most extreme effect estimates, and at the same time were also based on the smallest sample sizes. With their large standard errors, these two are still consistent with the remaining data (i.e., the intervals shown in the forest plot (Figure~\ref{fig:karner05}) are still mutually overlapping), and their estimates are shrunk considerably towards the remaining studies.
    
\begin{figure}
      \centering
      \makebox{\includegraphics[width=0.95\linewidth]{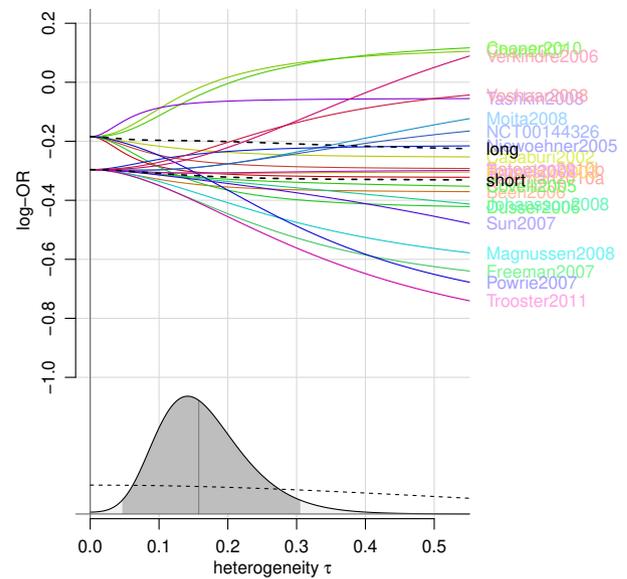}}
      \caption{\label{fig:karner02}Trace plot for the COPD example data.\citep{KarnerChongPoole2014} Meta-regression using a single binary covariable (``short'' vs. ``long'' follow-up).} 
    \end{figure}

    With some heterogeneity evident in the data, and a number of study characteristics recorded, it is interesting to investigate potential sources of heterogeneity. For example, studies were of differing duration (a case of \emph{methodological diversity}\cite{CochraneHandbook}), and it is quite conceivable that the treatment effect (OR) may differ for shorter or longer follow-up periods.\citep{RoeverAndreasFriede2016}
    A meta-regression including a binary covariable distinguishing ``short'' and ``long'' studies (defined as follow-up times up to $1$~year or more than $1$~year) is shown in Figure~\ref{fig:karner02}.
    Unlike in the previous example (Section~\ref{sec:app:MR1}), from the data it is not quite clear whether there actually exists a difference between the two groups, and the heterogeneity's posterior distribution is virtually unaffected by the inclusion of the covariable.

\begin{figure}
      \centering
      \makebox{\includegraphics[width=0.95\linewidth]{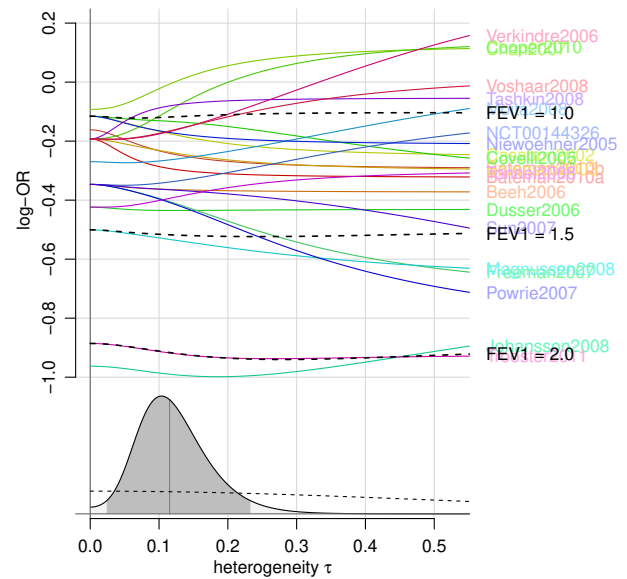}}
      \caption{\label{fig:karner03}Trace plot for the COPD example data.\citep{KarnerChongPoole2014} Meta-regression using a single continuous covariate (baseline FEV\textsubscript{1}).} 
    \end{figure}

    Another relevant determinant of the treatment effect may be the study participants' disease severity. A common measure of disease severity is  the \emph{forced expiratory volume in 1~second (FEV\textsubscript{1})}, which is determined through spirometry, and which quantifies a patient's breathing capacity. This amount is reduced with increasing COPD severity.\cite{Doherty2008} For the present data set, the population averages (at inclusion) are available for all $22$~studies.
    As this covariable relates to differences between study populations, this would be a case of \emph{clinical diversity}.\cite{CochraneHandbook}
    Figure~\ref{fig:karner03} shows a trace plot for the meta-regression considering the FEV\textsubscript{1} value as a covariable (via an \emph{intercept / slope} parametrization).
    
    On the left-hand side the effects are fully shrunken to the predicted effect size for the covariate value of that study.  There is a great deal of spread in these values, indicating that the covariate is important.  
    Many of the lines are flatter than in the other two plots, indicating less shrinkage being necessary than for the other models. Also, most of the lines are not as steep over the range of plausible values, indicating less sensitivity to the value of~$\tau$.
    Further, the posterior distribution of~$\tau$ has moved to smaller values, because more variability among studies is accounted for. The posterior median is at about~$0.12$, while the range of predicted values covers multiples of that, illustrating how a substantial fraction of heterogeneity is explained by the FEV\textsubscript{1} value.
    
    The traces for conditional effects at three selected FEV\textsubscript{1} values are also shown in the plot (for FEV\textsubscript{1} between~$1.0$ and~$2.0$, roughly covering the range of population means encountered in the data); one can see that larger FEV\textsubscript{1} values correspond to larger effects, suggesting that treatment benefit is greater in less severe cases.


\section{Frequentist trace plots}\label{sec:freqplot}
  Trace plots may also be motivated based on frequenstist reasoning. 
  The conditional distributions of effects (study-specific effects, overall effects or linear combinations of regression coefficients) have their analogues in so-called \emph{best linear unbiased prediction (BLUP)}. When considering uniform priors for effects ($\mu$ or $\beta_i$), the conditional posterior expectations and standard deviations correspond to frequentist conditional point estimates and standard errors.\cite{RaudenbushBryk1985,Viechtbauer2010,Robinson1991}
  While in a frequentist framework it is not possible to assign (prior or posterior) probabilities to heterogeneity~($\tau$) values, one may still mark confidence interval bounds or consider different $\tau$~values or ranges in the spirit of a sensitivity analysis.
  While there are similarities and analogies between the Bayesian and frequentist approaches, care needs to be taken regarding the differing interpretation, e.g., of credible and confidence intervals (in the Bayesian and frequentist contexts, respectively), but these issues are beyond the scope of the present investigation.\cite{AitkinLiu2018,MoreyEtAl2016}
  
  \begin{figure}
    \centering
    \makebox{\includegraphics[width=0.95\linewidth]{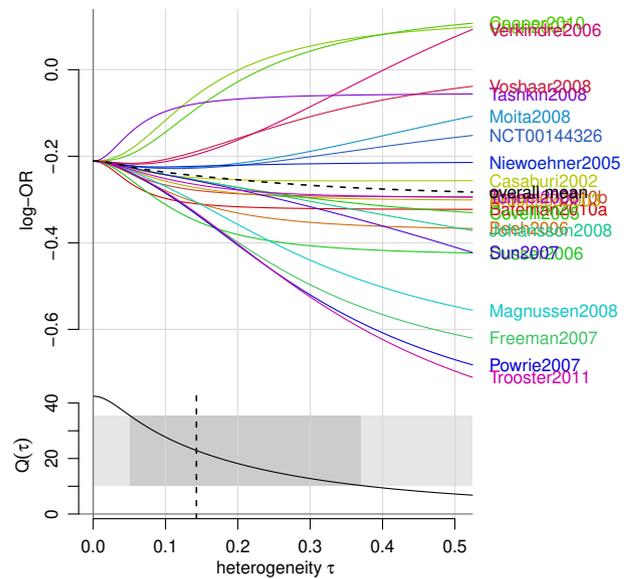}}
    \caption{\label{fig:karner04}Trace plot for the COPD example data\citep{KarnerChongPoole2014}, analo\-gous to Figure~\ref{fig:karner01}, but based on a frequentist analysis. The bottom panel illustrates the $Q$-test statistic as a function of the heterogeneity, and the resulting $Q$-profile confidence interval for~$\tau$; the dashed line indicates the point estimate for~$\tau$.}
  \end{figure}
  
  Figure~\ref{fig:karner04} shows a trace plot for the COPD example data (see also Figure~\ref{fig:karner01}) based on functions from the \texttt{metafor} package.\citep{Viechtbauer2005}
  The plot's bottom panel shows the $Q$-test statistic as a function of the heterogeneity~$\tau$ considered  as the (point) null hypothesis. The grey area indicates the central 95\% region based on a $\chi^2$~distribution with $(k-1)$ degrees of freedom (with $k$~denoting the number of studies); the points where the $Q$-statistic exceeds these bounds then constitute a confidence interval for~$\tau$, shown in dark grey. This bottom panel hence essentially illustrates the construction of a ($Q$-profile) confidence interval for~$\tau$.\cite{Viechtbauer2007} The dashed vertical line shows the heterogeneity point estimate (here: the \emph{restricted~ML} estimate $\hat{\tau}_\mathrm{REML}=0.14$). Analogous plots can be generated based on other test statistics, e.g., the likelihood ratio (or deviance).

  The plot highlights the differences between common frequentist and Bayesian treatments of the inference problem. In a Bayesian approach, effect estimates (for the overall mean~$\mu$ or for shrinkage estimates~$\theta_i$) result by averaging (marginalising) over the heterogeneity's posterior distribution (shown e.g. at the bottom of Figure~\ref{fig:karner01}). A common frequentist approach on the other hand is to treat the heterogeneity estimate~$\hat{\tau}$ as if this was known to be the true value (i.e., to \emph{condition on}~$\hat{\tau}$) and derive effect estimates by considering the corresponding vertical ``slice'' of the plot's top half. This is a reasonable approximation when $\tau$~is estimated with good accuracy, but otherwise it leads to overconfidence in the resulting effect estimates. Another approach at propagating heterogeneity uncertainty through to the effect estimates is by using adjusted standard errors and Student\mbox{-}$t$ quantiles as in the \emph{Hartung-Knapp-Sidik-Jonkman (HKSJ)} method.\citep{RoeverKnappFriede2015}


\section{Discussion}\label{sec:discussion}

  The trace plot is a little-used plot that conveys a great deal of useful information.  Usual methods of analysis, both fixed- (common-) and random-effect as well as empirical Bayes, commonly ignore uncertainty in the estimation of~$\tau$.  Bayesian methods take it into account, but average over values of~$\tau$, thus hiding the extra cause of variability.  The trace plot shows both the uncertainty in our knowledge of~$\tau$, but also the effect of that uncertainty on our knowledge of study effects and parameter estimates.  In addition, the plot allows us to see more clearly the presence of outliers or hidden subgroups of studies.  It is most useful for meta-analyses with small to moderate numbers of studies, because as the number of studies grows, our knowledge of~$\tau$ becomes stronger, and there is little variation in parameter or shrinkage estimates within reasonable ranges of estimates of $\tau$.  Furthermore, with many studies the trace lines can appear too tangled, making interpretation more difficult.

  If the trace lines are relatively flat over the area where the posterior of tau is appreciable, there is no sensitivity of the estimates to $\tau$.  If the lines are not flat, there is sensitivity to $\tau$.  Thus the plot gives valuable information about the sensitivity to the estimate of $\tau$.  This is especially important for likelihood or empirical Bayes meta-analysis, which rely on the estimate of $\tau$ to be precise for inference about the mean (or regression parameters).    

  While trace plots provide insights regarding the interplay of heterogeneity and effect and shrinkage estimates, they do not help much in judging the appropriateness of a prior. Choice of sensible priors depends on the scale of the endpoint and the context of the analysis, and varying prior shape or scale is part of sensitivity analyses in a Bayesian framework.\citep{RoeverEtAl2021,Roever2020} Varying priors will only affect the trace plot's bottom panel, while the top remains unaffected (besides possible changes in the range of $\tau$~values considered), as the effect estimates are \emph{conditional} on the heterogeneity.

  We can produce analogous plots from a frequentist perspective based on the best linear unbiased predictions (BLUP).  Figure~\ref{fig:karner04} was produced using results from the \texttt{metafor} package.\cite{Viechtbauer2010} 
  The bottom panel of the trace plot illustrates the inference on the heterogeneity, showing the point estimate as well as the $Q$-profile confidence interval along with the underlying $Q$-statistic.

  The trace plot is easy to produce in \texttt{bayesmeta} using the ``\texttt{traceplot()}'' function that is applied to the object returned by the analysis function (\texttt{bayesmeta()} or \texttt{bmr()}).\cite{bayesmeta,Roever2020,RoeverFriede2023}  
  For those who use \texttt{metafor},\cite{Viechtbauer2010} we provide code for that package in the Appendix.
  It also used to be available in the \texttt{hblm} package,\cite{hblm} but that has never been officially released for~\textsf{R}, and the \textsc{S-plus} version is no longer on the web, nor would it work in~\textsf{R} without modification.

  The general ideas underlying the trace plot should be generalizable beyond the normal-normal hierarchical model (NNHM) discussed here. 
  An obvious example would be network meta-analyses (NMAs); as long as these may be expressed as special cases of a simple meta-regression (e.g., when only pairwise comparisons are included), these would be tractable using the tools shown in Section~\ref{sec:applications} already.\cite{RoeverFriede2023}
  In principle, as long as there is only a single heterogeneity parameter involved, the same approaches should generally still work for NMAs, and these might be straightforward to implement, e.g., based on existing functions in the \texttt{netmeta} \textsf{R}~package.\cite{BalduzziEtAl2023}
  However, while computations are straghtforward (and mostly analytical) in ``simple'' normal models, technical calculations would be more demanding, e.g., in the case of a binomial-normal model; and these might in fact be easier in a frequentist framework compared to a Bayesian one.


\section*{Acknowledgment}
Support from the \emph{Deutsche Forschungsgemeinschaft (DFG)} is gratefully acknowledged (grant number \mbox{FR~3070/3-1}).

\section*{Conflicts of interest}
  The authors have declared no conflict of interest.

\section*{Data availability}
  The data and code that supports the findings of this study are available in the supplemental material of this article.
  The data sets discussed are available from the cited references, but are also included in \textsf{R}~packages; the SAT-coaching, Aspirin and COPD data are available in the \texttt{bayesmeta} \textsf{R}~package,\cite{bayesmeta} and the {\notwo}~data are available in the \texttt{metadat} \textsf{R}~package.\cite{R:metadat}

\section*{Highlights}
\paragraph{What is already known:}
\begin{itemize}
  \item heterogeneity is a critical aspect in the interpretation of a meta-analysis
  \item a number of plots have been proposed to illustrate and diagnose the results of meta-analyses and meta-regressions (such as forest plots, L'Abbe plots, \ldots).
\end{itemize}

\paragraph{What is new:}
\begin{itemize}
  \item the \emph{trace plot} has been proposed (originally in the 1980s already) to illustrate the interplay of effect estimates and (estimated) heterogeneity.
  \item it is useful in the context of (``simple'') meta-analyses or meta-regressions, and may be derived in the context of Bayesian as well as frequentist approaches.
\end{itemize}

\paragraph{Potential impact for readers outside the authors’ field:}
\begin{itemize}
  \item similar plots illustrating (conditional) shrinkage effects might be useful in other contexts, as well as in more general settings.
\end{itemize}

\bibliography{literature}

\clearpage


  \appendix \label{sec:appendix}
  
  \section{Trace plot for SAT coaching data with credible intervals}
\begin{figure}
      \centering
      \makebox{\includegraphics[width=0.95\linewidth]{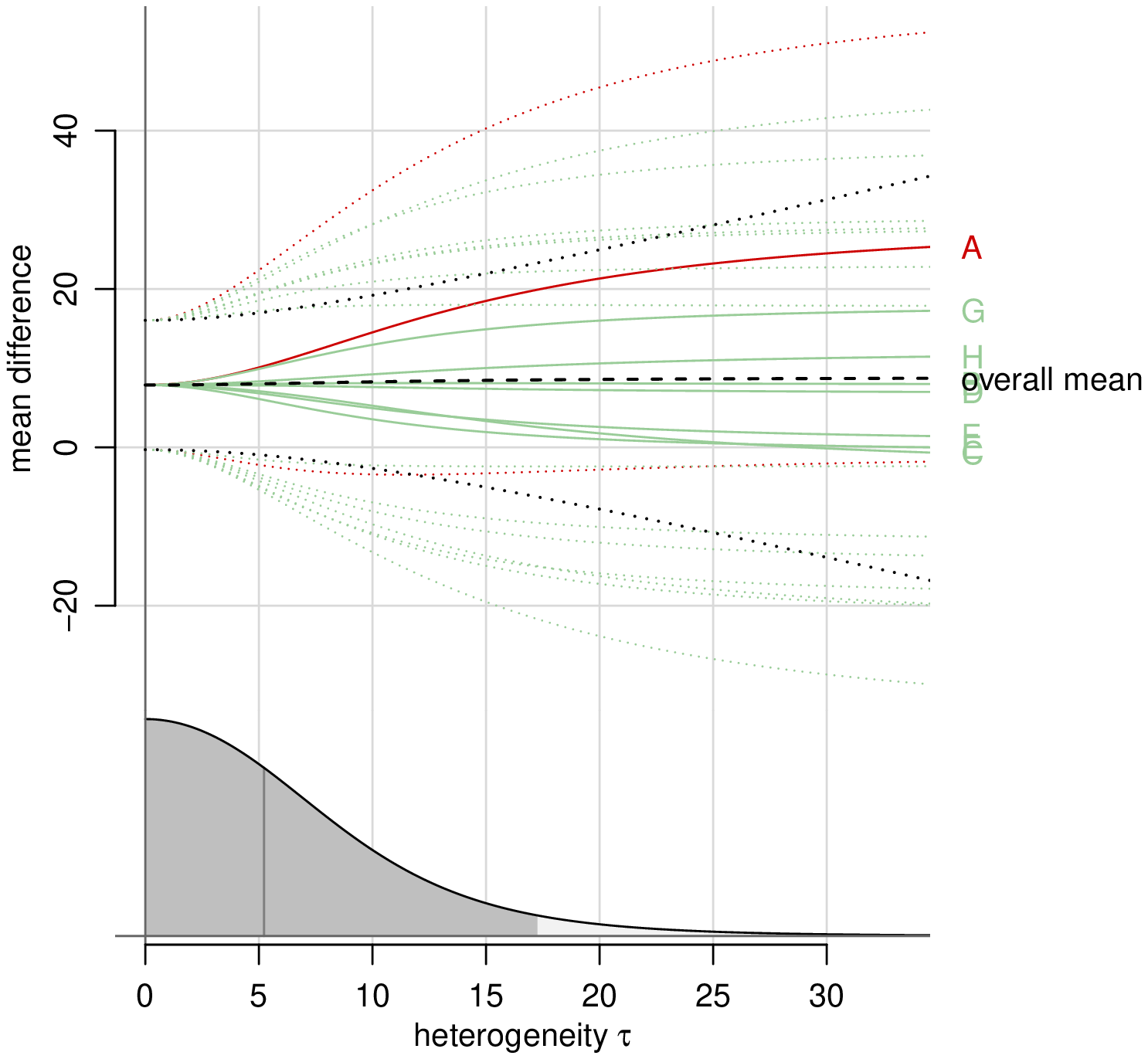}}
      \caption{\label{fig:rubin06}Trace plot for the SAT coaching example data (see Section~\ref{sec:app:MA1}) including (conditional) credible ranges. As it is very busy plot, only the estimates and CIs for \emph{school~A} (red) and for the overall mean (black) are highlighted, the remaining 7~estimates are shown in pale green. Conditional means are shown as solid lines, 95\%credible limits are shown as dotted lines.\citep{Rubin1981}}
    \end{figure}

    Figure~\ref{fig:rubin06} shows a trace plot for the SAT coaching example data from Section~\ref{sec:app:MA1};\cite{Rubin1981} the plot is analogous to Figure~\ref{fig:rubin02}), but includes (conditional) credible bounds.
  
  \section{Forest plot for SAT coaching data with estimates}
    \begin{figure}
      \centering
      \makebox{\includegraphics[width=0.95\linewidth]{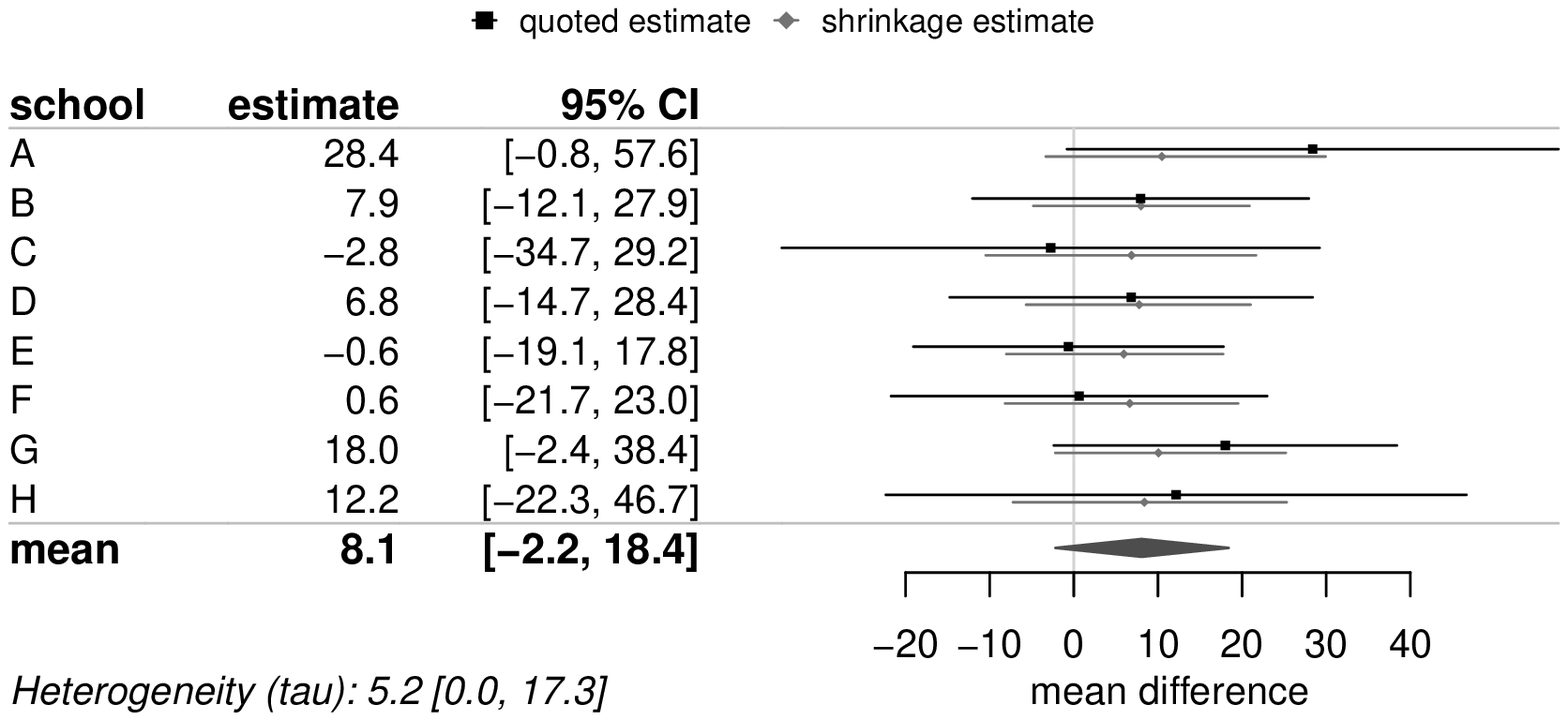}}
      \caption{\label{fig:rubin05}Forest plot for the SAT coaching example data, including overall mean and (marginal) shrinkage estimates.\citep{Rubin1981}} 
    \end{figure}
    Figure~\ref{fig:rubin05} shows the data from Section~\ref{sec:app:MA1} in a forest plot along with shrinkages estimates ($\theta_i$), overall mean~($\mu$) and prediction~($\theta^\ast$) based on an analysis with uniform priors. The eventual (marginal) shrinkage estimates result from averaging the \emph{conditional} shrinkage estimates over the heterogeneity's posterior distribution.
  
  \section{Rubin's 1981 plot}\label{sec:RubinAppendix}
\begin{figure}[h]
      \centering
      \makebox{\includegraphics[width=0.90\linewidth]{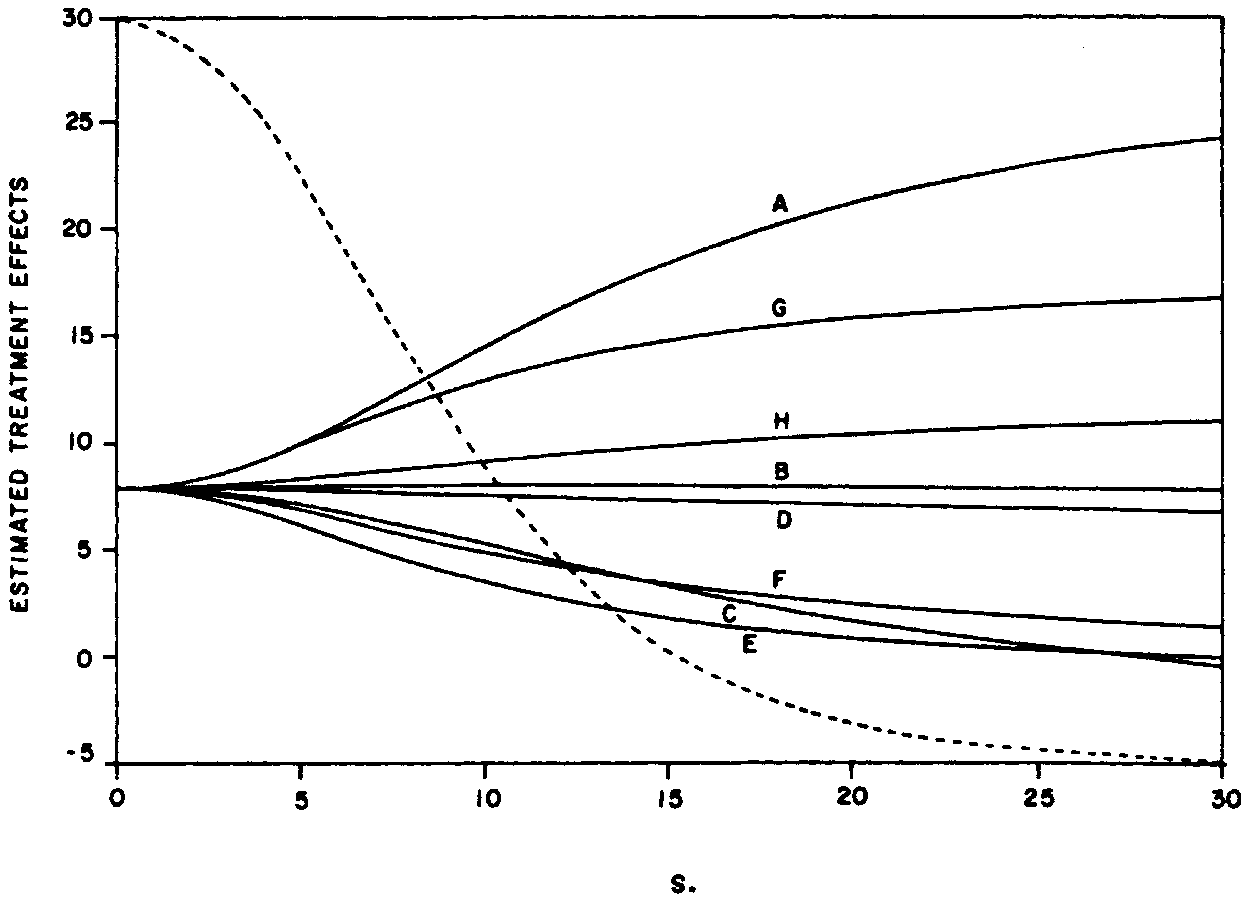}}
      \caption{\label{fig:rubin04}Trace plot for the SAT coaching example data from the original publication (Rubin, 1981)\citep{Rubin1981} (Figure caption:\emph{``Estimates of treatment effects as functions of~$S_\ast$ with likelihood of~$S_\ast$ superimposed.''}, where $S_\ast$~corresponds to the heterogeneity standard deviation~$\tau$ here).} 
    \end{figure}

    Figure~\ref{fig:rubin04} shows a trace plot for the SAT coaching example data from Section~\ref{sec:app:MA1} as shown in Rubin's original publication.\cite{Rubin1981}

  \section{The hblm plot}\label{sec:hblmAppendix}
\begin{figure}[h]
      \centering
      \makebox{\includegraphics[width=0.90\linewidth]{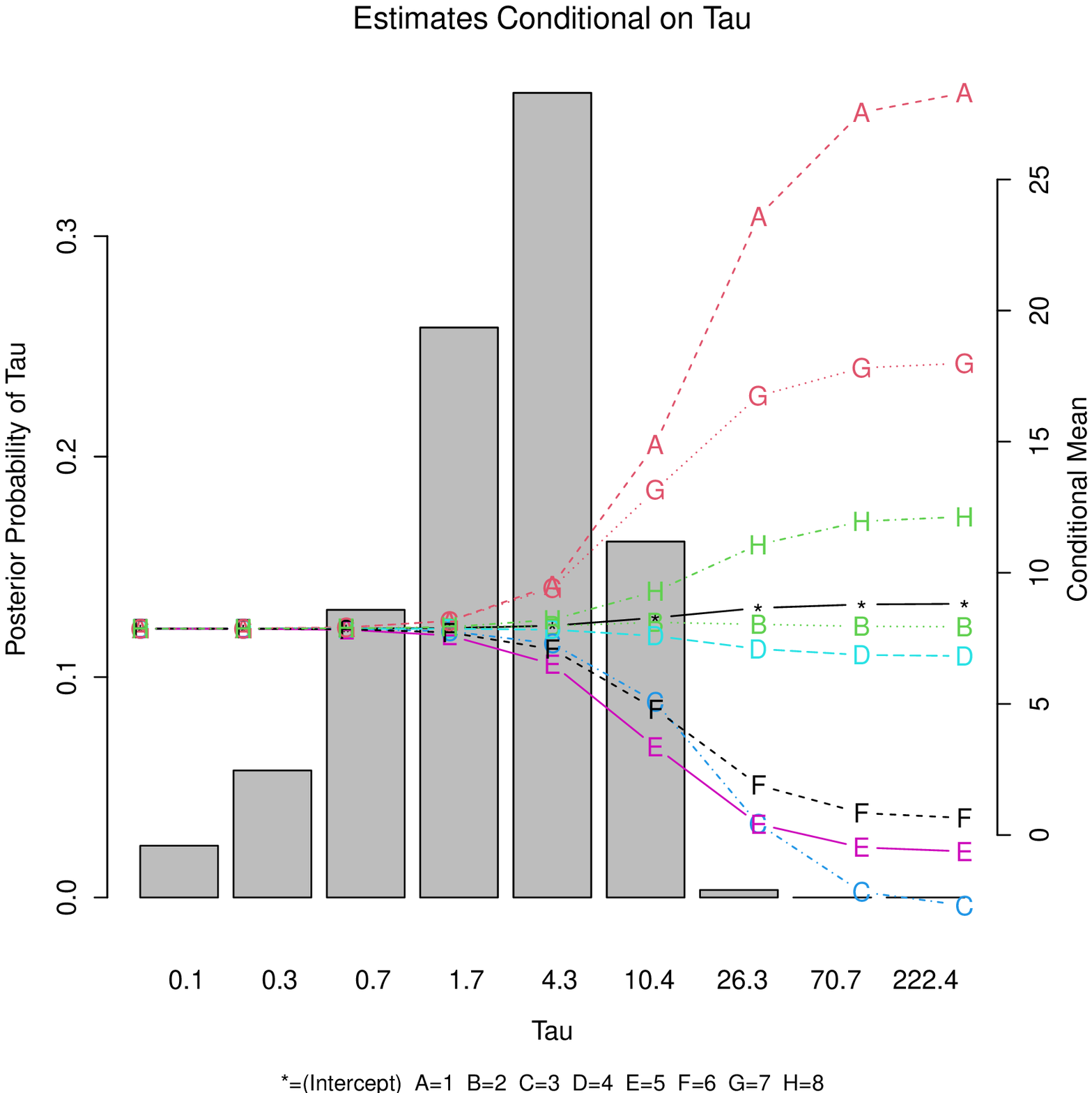}}
      \caption{\label{fig:rubin03}Trace plot for the SAT coaching example data (generated using the \texttt{hblm}~package).\citep{DuMouchel1994,hblm} Note the scaling of the $x$-axis.} 
    \end{figure}

    Figure~\ref{fig:rubin03} shows a trace plot for the SAT coaching example data from Section~\ref{sec:app:MA1} as generated by the \texttt{hblm} \textsc{S-plus} package (actually: a code version ported to~\textsf{R}). 
    Note the odd scaling of the (discretized)  $x$-axis, and the very different appearance compared to Figure~\ref{fig:rubin02}.

  \section{Forest plots for example data}\label{sec:ForestAppendix}
\begin{figure}
      \centering
      \makebox{\includegraphics[width=0.95\linewidth]{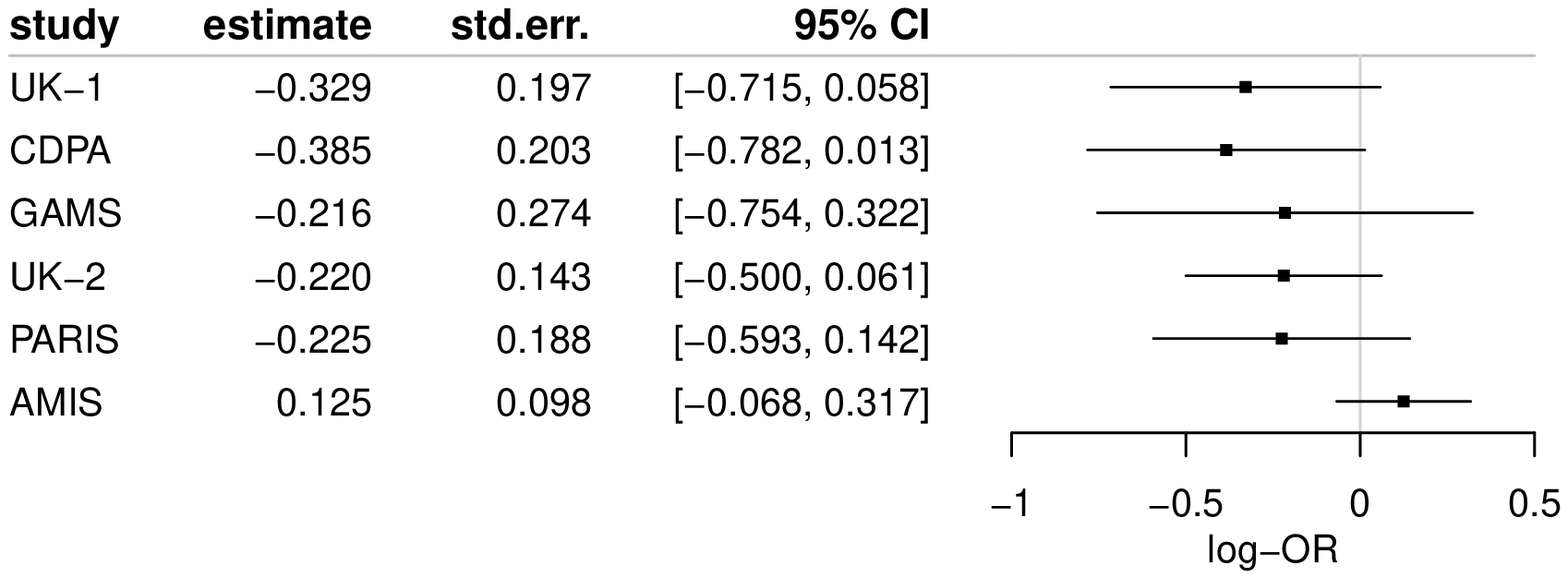}}
      \caption{\label{fig:peto02}Forest plot for the Aspirin example data.\citep{Peto1980}} 
    \end{figure}

    Figure~\ref{fig:peto02} illustrates the data of the Aspirin example from Section~\ref{sec:app:MA2}.
    
\begin{figure}
      \centering
      \makebox{\includegraphics[width=0.95\linewidth]{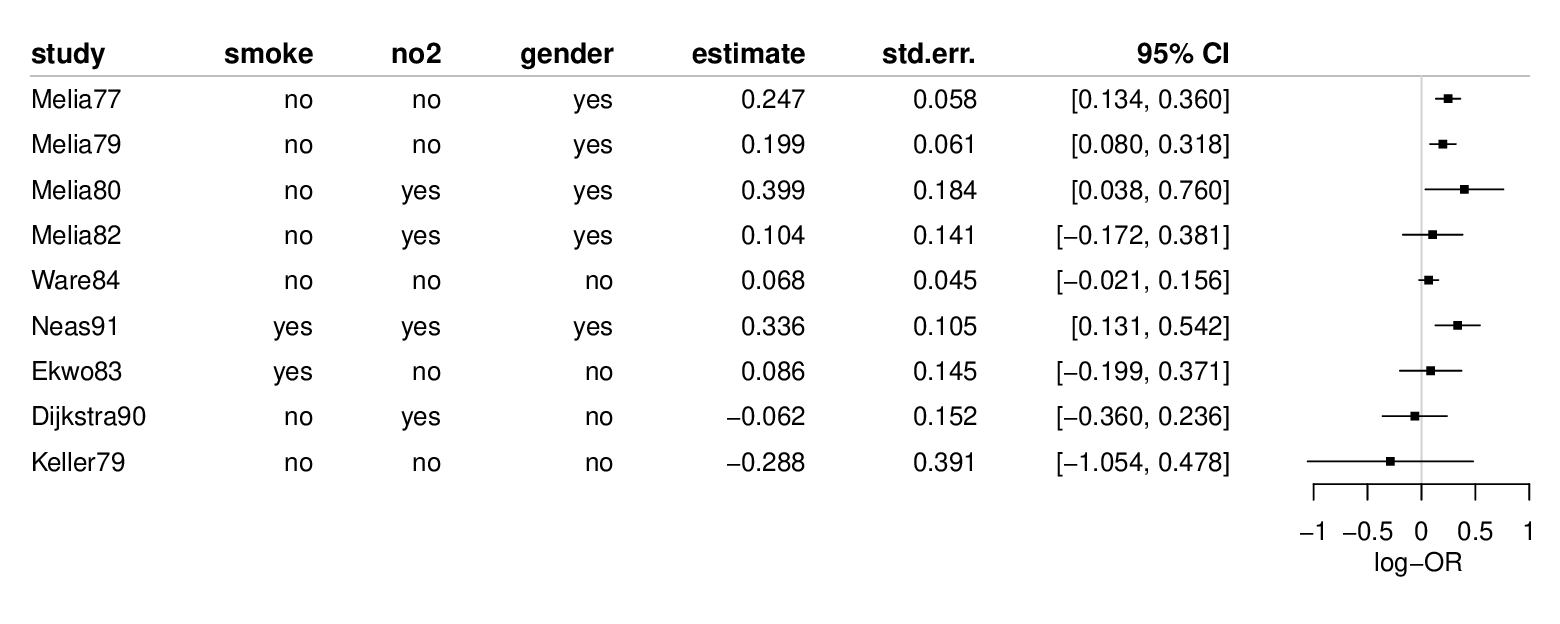}}
      \caption{\label{fig:dumouchel04}Forest plot for the {\notwo} example data.\citep{DuMouchel1994}} 
    \end{figure}

    Figure~\ref{fig:dumouchel04} illustrates the data of the {\notwo} example from Section~\ref{sec:app:MR1}.
    
\begin{figure}
      \centering
      \makebox{\includegraphics[width=0.95\linewidth]{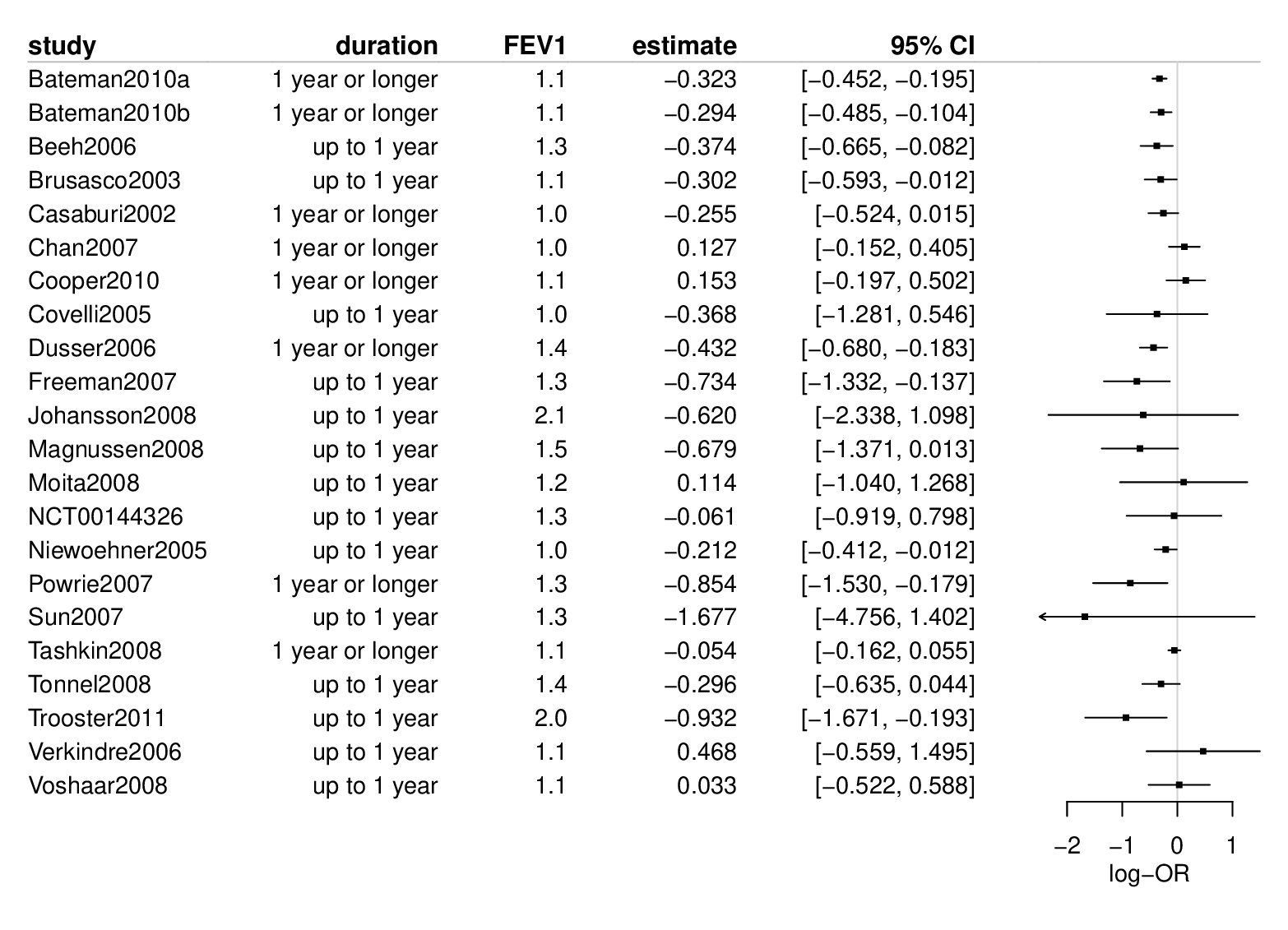}}
      \caption{\label{fig:karner05}Forest plot for the COPD example data.\citep{KarnerChongPoole2014}} 
    \end{figure}

    Figure~\ref{fig:karner05} illustrates the data of the COPD example from Section~\ref{sec:app:MR2}.

  \section{Example \textsf{R}~code}
    \textsf{R}~code to reproduce a simple trace plot (Fig.~\ref{fig:rubin02}), using the \texttt{bayesmeta} \textsf{R}~package:
\begin{verbatim}
# load "bayesmeta" R package:
library("bayesmeta")

# load example data:
data("Rubin1981")

# perform meta-analysis (uniform priors):
bm <- bayesmeta(y=Rubin1981[,"effect"], 
                sigma=Rubin1981[,"stderr"],
                labels=Rubin1981[,"school"], 
                tau.prior="uniform")

# show meta-analysis results (Fig. B2):
forestplot(bm)

# show trace plot (Fig. 3):
traceplot(bm)
\end{verbatim}

\end{document}